\documentclass{llncs}

\usepackage{mathrsfs}

\usepackage{url}

\usepackage{amssymb}
\usepackage{stfloats}

\usepackage{graphicx}

\usepackage{psfig}
\usepackage{bm}

\usepackage{times}
\usepackage{helvet}
\usepackage{courier}
\usepackage{url}

\usepackage{float}
\usepackage{subfigure}
\usepackage{subfig}
\usepackage{caption}

\usepackage{url}
\usepackage{booktabs}
\usepackage{multirow}

\usepackage{epstopdf}

\usepackage{CJK}
 \usepackage{amsmath}

\author{Chang Liao}
\institute{Dongguan Securities Company Limited,  China }

\begin{document}

\title{Prospecting Community Development Strength based on Economic\\ Graph: From Categorization to Scoring}

\titlerunning{Prospecting Community Strength}

\maketitle

\begin{abstract}
Recent years have witnessed a growing number of researches on community characterization. In contrast to the large body of researches on the categorical measures (rise or decline) for evaluating the community development, we propose to estimate the community development strength  (to which degree the rise or decline is). More specifically,   given already known categorical information of community development, we are attempting to quantify the community development strength, which is of great interest.  For example, considering the task of predicting the development of the employee size for a given company, conventional work  only predicts whether the employee size will grow or not. Our setting, on the other hand, can learn a score making reflection on the growth of employee size. The scoring task fails to get a preferable solution by learning an attribute-based classifier, as the competitive advantage of external  relations in community strength has not been fully assessed. Such relation can facilitate productive activity, and  we call the relation as  impact asset. Motivated by the increasing availability of large-scale data on the network between entities among communities, we investigate how to score the the community's development strength. We formally define our task as prospecting community development strength from categorization based on multi-relational network information and identify two challenges as follows: (1) limited guidance for integrating entity multi-relational network in quantifying the community development strength; (2) the existence of selection effect that the community development strength has on network formation. Aiming at these challenges,
we start by a hybrid of discriminative and generative approaches on multi-relational network-based community development strength quantification. Then a network generation process is exploited to debias the selection process. In the end, we empirically evaluate the proposed model by applying it to quantify enterprise business development strength. Experimental results demonstrate the effectiveness of the proposed method.

\end{abstract}

\section{Introduction}

In recent years, remarkable efforts have been devoted to characterizing/ profiling communities, such as  \cite{Zheng2017From,LinZZZWX17}. Different from community detection task, community characterization/ profiling  aims to give a description of the community's fundamental characteristics. Among them, predicting community development is of great interest, especially in marketing and economics fields.  For instance, venture investors can get better insights into a company's quality based on more precise prediction of its development. By trading on such information, they may achieve lucrative extra returns.

However, most of these predictions from analysts are in the form of categorical rating (binary value, rising or declining), and little is known  about the degree of the increase or decrease of  development. When such quantitative information is available, it enables investors to be better informed of the value differences between companies. Take Facebook as an example: considering the recent information leakage crisis for Facebook, although investors can intuitively deduce this incident has a negative impact on the development of the company, it is hard for them to predict the extent of this impact. It is naturally an unsupervised task setting for measuring community development strength, where the weighing parameter for each feature factor cannot be well determined, leading to uncontrolled results. Faced with real-world application scenario,  often categorical information for each community is publicly available. Under such setting, it turns out to be a weak-supervised task and can overcome the critical issues that unsupervised setting poses. For instance, companies can be classified to different performance categories by analysts. And our task is not to predict the development category (rising or declining), but quantify the fined grained development strength for each company given the performance categorical information.

\begin{figure}[t]

\centering

 \includegraphics[width=.982345683020178521789893019824\textwidth]{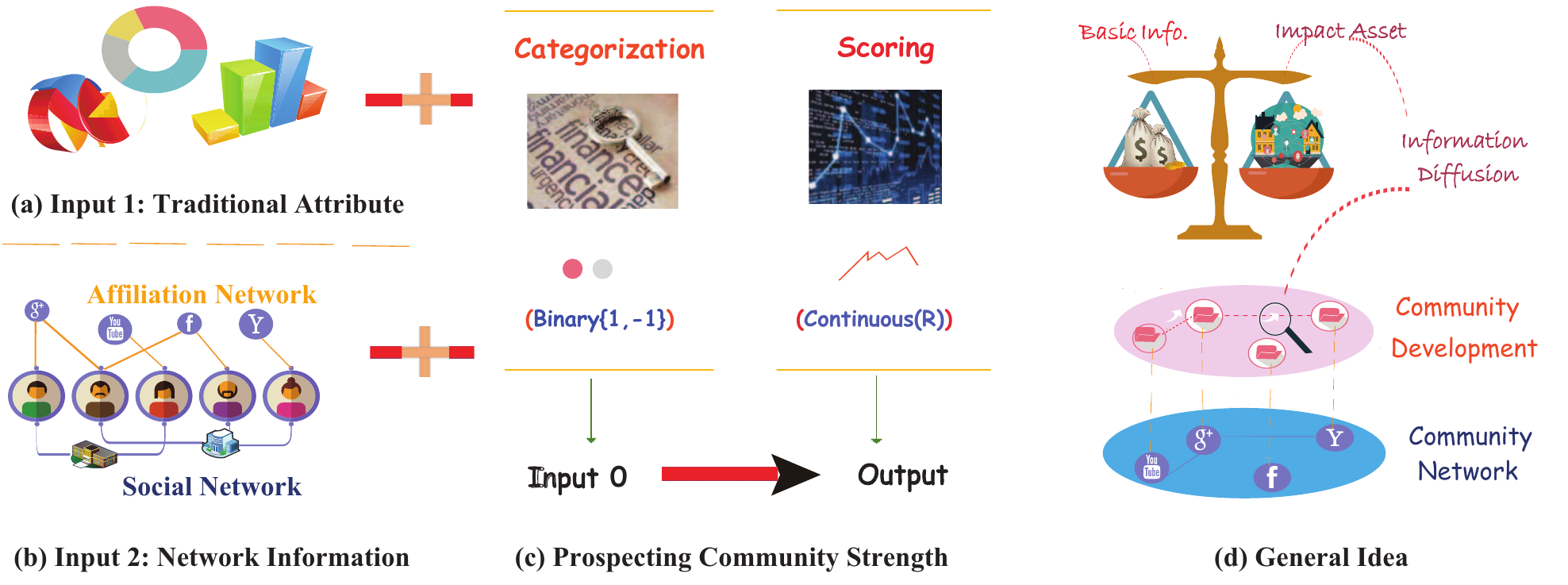}

\caption{An overview of  prospecting community development strength. The proposed task is to reconstruct a score instead of a flat categorical value by integrating  traditional financial attributes and network information. The general idea  is to model network information as impact asset, where new dataset of both affiliation (such as directors and their companies) and social network (such as friends on twitter or members in the same club ) are leveraged.}

\label{fig:1}

\end{figure}

A straightforward method to quantify the community's development strength from categorical information is to learn an attribute based classifier and treat the value of loss function as the strength, as is shown in Figure \ref{fig:1}(a).  However,  such information may not be enough for development strength quantification.  That is, the development of a community is also affected by  external relations.
Intuitively, it is easy for external relations to interfere with (improve or decrease) the quantitative scoring to some degree. More specifically, network relations can play a crucial role in a variety of mechanisms,   from information diffusion to
the adoption of political opinions and technologies (\cite{Marlow2012The}). As in corporate finance, a  company's development strength can also be influenced by its impact asset (\cite{GlobalImpact}), the interpersonal relations between this company and others.
For example, corporations can get lower interest rates if their board members have strong social connections with banks. By studying the market reaction to the departure of independent directors, \cite{nguyen2010the} shows
that outside directors account for an average of 1\% of the firm value.  Such relations can facilitate productive activity.

Fortunately, there is more and more online information available, such as  large-scale data on the information of entities in communities, which  has the potential to reveal the specific value/ capital of them. For instance, as  is shown in Figure \ref{fig:1}(b), in social networks, entities (employees) are linked to each other by multi-type relations; meanwhile, entities often coexist in a two-mode affiliation network, in which entities are linked to communities (companies). This brings the opportunity to quantify community development strength based on network information between entities among organizations. To highlight  the  economic value of  network information, we  define  a collection of  networks (entity - organization affiliation network and entity-entity social network)  with rich attributes as an economic graph. To fill the gap of the lack of researches on quantifying community strength,  as well as motivated by the available entity network information, we propose to study a new problem - prospecting community development strength from observed categorical information to scoring based on economic graph, as in Figure \ref{fig:1}(c).

\begin{figure}[t]

\centering

%
%
\subfigure[Basic Model]{%
  \includegraphics[width=.2345683020178521789893019824\textwidth,height=0.168301245783120286851243048402363064022147\textwidth]{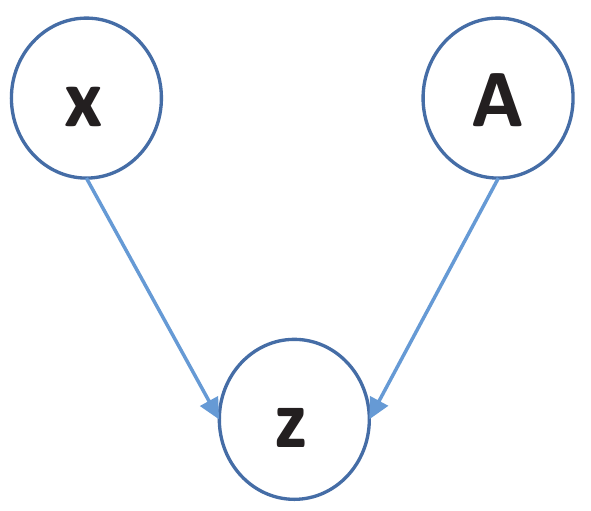}   }
\subfigure[Multi-relational Network]{%
  \includegraphics[width=.301363024318504205304025830243\textwidth,height=0.168301245783120286851243048402363064022147\textwidth]{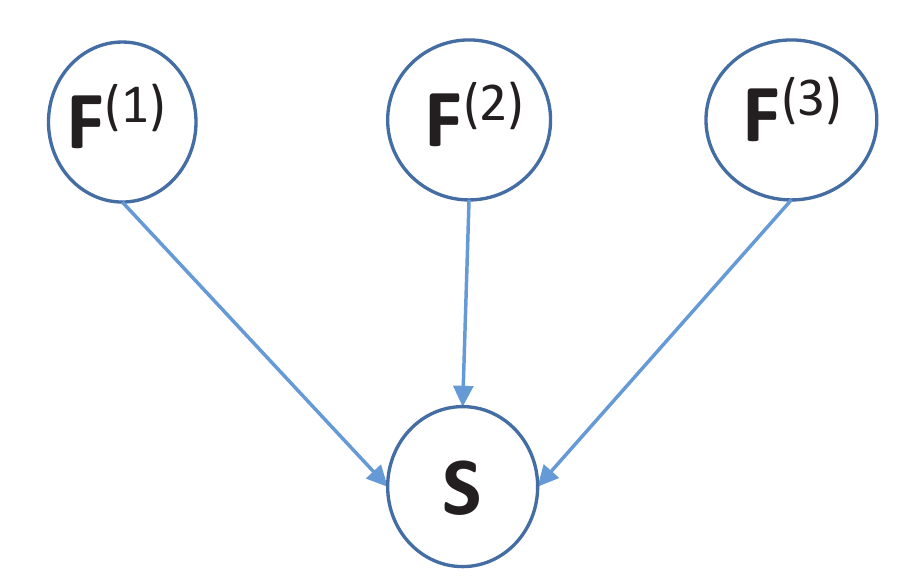} }
\subfigure[Selection Bias]{%
  \includegraphics[width=.20345683020178521789893019824\textwidth,height=0.168301245783120286851243048402363064022147\textwidth]{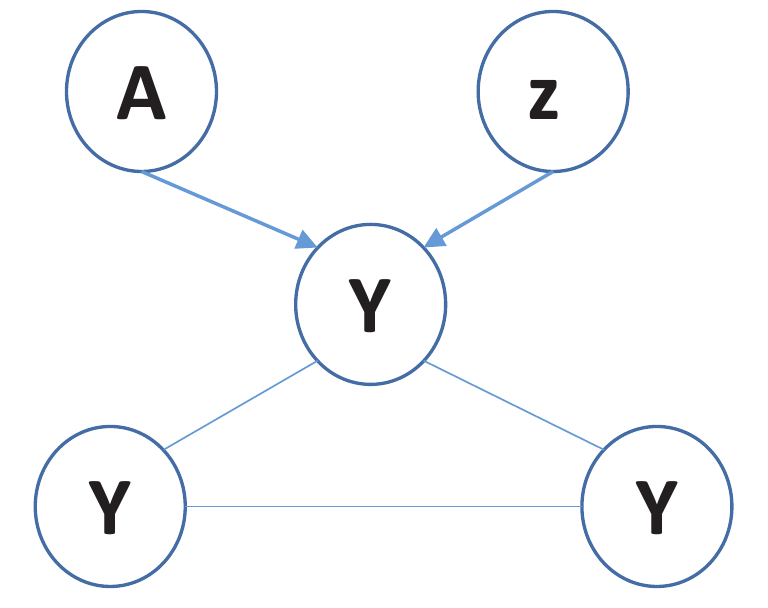}   }

\vspace{-0.2864212485232cm}

\caption{Challenge and Solution Illustrations, where vector $\mathbf{x}$ is community development categorical information, matrix $\mathbf{A}$ is community attribute, matrix $\mathbf{Y}$ is network edge and vector $\mathbf{z}$ is  community development score, tensor $\mathbf{F}$ is multi-relational social network and matrix $\mathbf{S}$ is weighted social network. In (a),  development quantification is based on attribute from categorization information. In (b), multi-relational links are integrated and weighted into single network. In (c),  community development strength and attributes contribute to network formation.}

\label{fig:3}

\end{figure}

To undertake the aforementioned task, we are faced with the challenge of limited guidance for integrating entity network for the community development strength quantification.  Compared with traditional categorization where standard classification methods can be applied, as in Figure \ref{fig:3}(a), our setting is totally different.  Characterizing network information as impact  and integrating it for community development strength quantification is difficult. Moreover, we are faced with is different importance of multi-relational links on community's development. Although  there is some empirical work on investigating the effect of affiliation network on certain activities from economic research, the entity  social network remains largely unexplored. For example, employees  can receive information from their social network neighbors via different channels/links. Different links may have different levels of importance, and the problem lies in how to integrate and weigh heterogeneous link information, as is shown in Figure \ref{fig:3}(b). In response, a hybrid  of discriminative and generative model  is proposed  for  network based development strength quantification. We show this model converges to a unique equilibrium under weak assumptions. Meanwhile, multi-relational links are combined  in development strength quantification by learning weights effectively.

Another major challenge is  the  selection effect that the community development strength has on network formation. While  the development of a community can be affected by its affiliation  network, its development strength can also affect affiliation network formation. This is called selection bias, which is illustrated in Figure \ref{fig:3}(c).  As an example of  board relation formation and firm development,  corporations would carefully  choose their boards to maximize their profit. There is a need to correct such selection bias, without which the network influence might be amplified. Various applications that rely on true network value  will not work effectively under biased settings. To tackle this challenge, network formation is modelled to mitigate possible selection bias.

This work contributes to a new problem of quantifying community development strength from categorization to scoring.  A collection of  networks (entity - organization affiliation network and entity - entity social network)  with rich attributes is refined as an economic graph and utilized,  which has not been addressed much in data mining research literature.

From technical perspectives,  we have identified three challenges. To tackle the aforementioned issues, we introduce a framework of Graph based community development strength quantification. Meanwhile, multi-relational links are combined  in development strength quantification by learning weights effectively.   A hybrid  of discriminative and generative model  is proposed. Network formation is jointly modelled to mitigate possible selection bias.

From application contribution perspectives, our model can be of great value in  business applications. In particular, we apply it to a corporate finance application scenery - measuring enterprise business development strength.  To the best of our knowledge, this is the first paper that studies the problem of quantifying company performance (development) strength based on  both social network and affiliation network of directors.

%
%
%
%
%
%
%

\section{Problem Definition}

In this section, we first introduce several concepts and then formally formulate the problem. Table 1 shows the important notations we will use in this paper.


\begin{small}

\begin{table}[h]
\caption{Several important mathematical notations}

\centering
\begin{tabular}{c|c|c|c}
\hline
 Notat. & Meaning & Notat. & Meaning\\
\hline
        $\mathbf{G}$/ $\mathbf{A}$ & Economic Network/  Attributes  & $\mathbf{Y}$/ $\mathbf{H}$ & Association/ Affiliation Network    \\

          $\mathbf{x}$/ $\mathbf{z}$  &  Development Categorization/ Scoring &  $\mathbf{S}$/ $\mathbf{F}$ & Social Network Matrix/ Tensor \\

          $\alpha$, $\mathbf{\beta}$, $\lambda$   &  Community Development Parameters &     $w$, $\mathbf{\eta}$, $\rho$ &  Network Formation Parameters   \\

          $u$ / $\psi$ & Utility Function  / Strength Difference  & $\mathbf{\xi}$/ $q$ & Weighing Parameters/ link type \\

        $\mathbf{\varepsilon}$/ $\mathbf{\overrightarrow{\varepsilon}}$ & Error Vector  &  $\mathbf{l}$ / $\mathbf{I}$ & Vector with Ones / Identity Matrix\\

        $\mathbf{\varphi}$ & Network Statistics   &  $a$, $b$ & Discriminative function parameters\\

          $\sigma_\varepsilon^2$/ $\sigma_{{\overrightarrow{\varepsilon}}}^2$ & Variance Term &  $\mathbf{\overrightarrow{z}}$ & Robust Development Score\\

\hline
\end{tabular}
\end{table}

\end{small}

\newtheorem{dep1}{Definition}
\begin{dep1}[\textbf{Economic Graph}]
Given $n$ communities and $m$ entities, an economic graph  is formally defined as ${\mathbf{G}}(\mathbf{F}, \mathbf{H}, \mathbf{A})$, which consists of
three components: a social graph between entities denoted  as $\mathbf{F} \in {\mathbb{R}^{ m \times m \times q}} $ with $q$ as the number of link types, an affiliation network between entities and communities as  $\mathbf{H} \in {\mathbb{R}^{ n \times m }} $, and generalized attributes  $\mathbf{A} \in {\mathbb{R}^{ n \times p}}$ with features of the community itself and its affiliation entities are included of $p$ as the attribute dimension.

\end{dep1}
To well understand the economic graph, we show it with an example about firms in Figure 3.
There are typically three parts:  (1) entity social network-there are multiple types of social relations between directors;  (2) entity affiliation network-a tie exists between a firm  and a director  if the director is a member of the firm; (3) aggregated attributes-both the firm's attribute and its directors' attribute information are combined. The economic graph  can express the data in various application area, such as
forums with entities comprised of different types of  link information, companies with directors consisting of heterogeneous relations.
In order to apply the economic graph to community-oriented organizations, two networks are transformed from entity (blank matrix) to community (shadow matrix) representation, as shown in Figure 3: (1) community association network transformed by entity affiliation network; (2) community social network transformed by entity social network and entity affiliation network.
\newtheorem{dep1213}[dep1]{Definition}
\begin{dep1213}[\textbf{Prospecting Community Development Strength}]

Given an economic graph $G =  \langle {\mathbf{F}, \mathbf{H}, \mathbf{A}} \rangle$ and categorical information $\mathbf{x} \in {\{0,1\}^{ n}}$,  our goal is to quantify community development strength $\mathbf{z} \in {\mathbb{R}^{ n}}$. The development score $\mathbf{\overrightarrow{z}}$ is to represent the full spectrum of development strength from weak to strong, in contrast to a coarse representation
of flat categorical value $\mathbf{x}$.

\end{dep1213}

A community of good development score should meet two criteria: (1) financial attributes-it has good financial attributes; (2) social capital-it has intensive connections with other communities of good development strength. The two criteria demonstrate the necessity of  the three elements of the economic graph $\mathbf{G} =  \langle {\mathbf{F}, \mathbf{H}, \mathbf{A}} \rangle$. In this paper, we aim to quantitatively evaluate the development strength  of the communities. More concretely, we attempt to recover a  score $\mathbf{\overrightarrow{z}}$ of the community development by utilizing economic graph under weak-supervision settings (flat categorical value $\mathbf{x}$). Compared with traditional settings, this paper proposes a new problem of community characterization - prospecting community development strength based on its categorization by utilizing economic network information.
\begin{figure}[t]
\centering


  \includegraphics[width=.88904749846821598767820624\textwidth, height=0.40327597324840362204683162028423\textwidth]{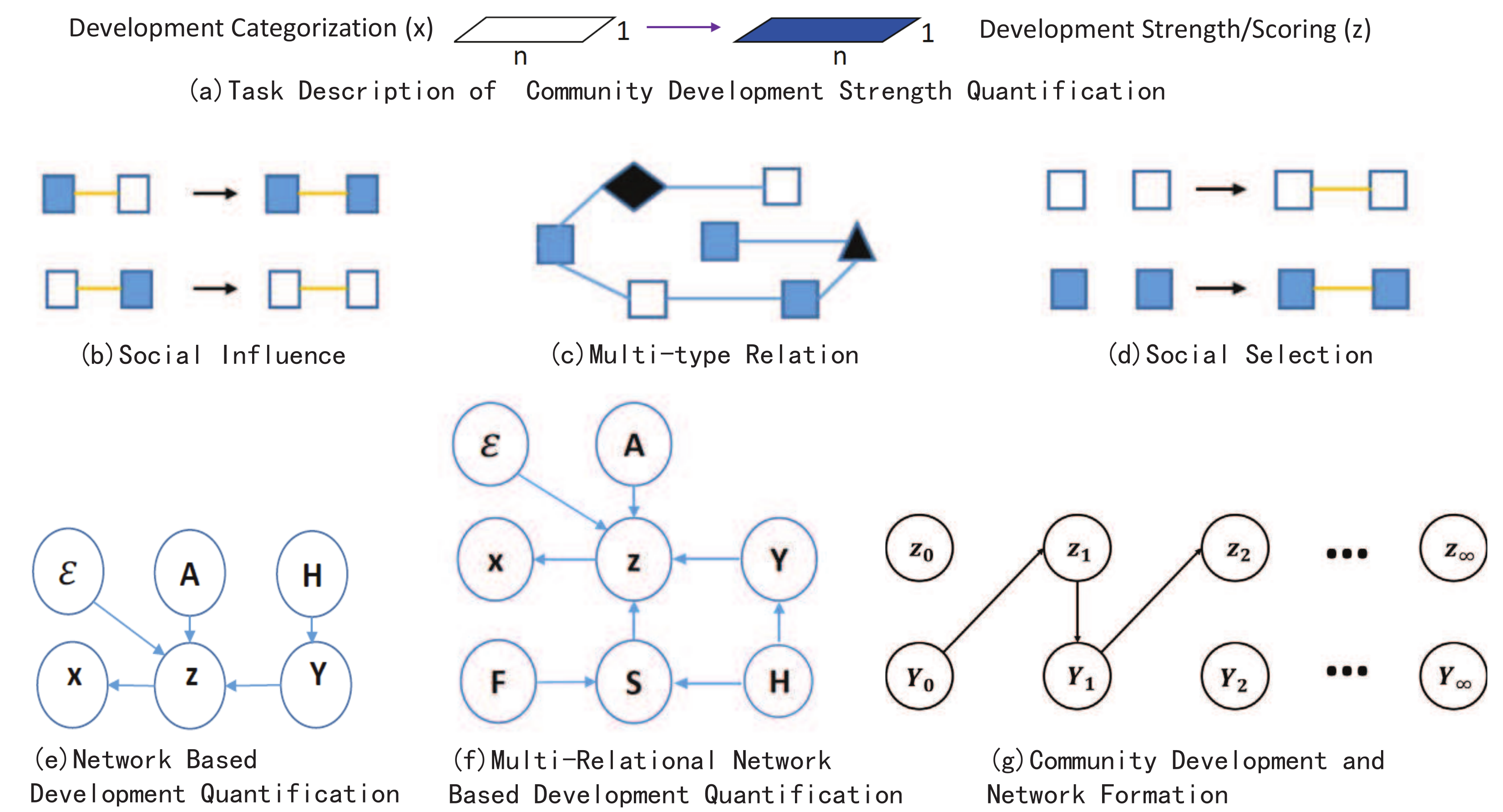}

\caption{Model Overview : (1) the studied  problem  in (a);  (2) model for network based development quantification in (e), improved model by integrating multi-relational links in (f),  improved model by considering network selection in (g); (3) intuitions for corresponding models in (b), (c), (d). Table 1 shows the meanings of these notations. }

\label{fig:4}
\end{figure}
%
%
%
%
%
%
%
%
%
%
%
%
%
%
%
%
%
%
%
%
%
%
%
%
%
%
%
%
%
%
%
%
%
%
%
%
%
%
%
%
%
%
%
%
%
%
%
%
%
%
\section{Framework Overview}

 And then, we propose a general framework  to make use of economic graph to quantify community development strength based on categorization, as shown in Figure \ref{fig:4}(a). The main process consists of a hybrid of discriminative approach (network based development strength quantification) and generative approach (categorical value generation based on development score), as in Figure \ref{fig:4}(b). This main process is joined with  multi-relational links integration,  as shown in Figure \ref{fig:4}(f), together with network formation process for selection bias mitigating, as shown in in Figure \ref{fig:4}(g).

\subsection{Network Based Development Strength Quantification}

Social influence among economic network refers to community development is affected by its neighbors. Motivated by this,  we relate economic network ${\mathbf{G}}(\mathbf{F}, \mathbf{H}, \mathbf{A})$ and community development categorization $\mathbf{x}$ to quantify community development strength $\mathbf{z}^*$, As in Figure 4(e). We assume that the development score is the effect
of economic network and the hidden determinant of the categorical information. This naturally leads to
a latent variable model, and the joint  probability can be written as

\begin{small}
\begin{equation}\label{equ:1}
\begin{array}{l}
 \Pr(\mathbf{H},  \mathbf{A}, \mathbf{x}; \mathbf{z}^*, \mathbf{Y}, \varepsilon) \propto \Pr(\mathbf{Y}|\mathbf{H}) \cdot  \Pr(\mathbf{z}^*| \mathbf{A},  \mathbf{Y}, \lambda, \alpha, \mathbf{\beta}, \varepsilon) \cdot \Pr(\mathbf{x}| \mathbf{z}^*, a, b) \cdot  \Pr(\varepsilon),
  \end{array}
\end{equation}
\end{small}

where  $\Pr(\mathbf{Y}|\mathbf{H})$ denotes the probability of community association network transformation from entity-community affiliation network,  and $\Pr(\mathbf{z}^*| \mathbf{A},  \mathbf{Y}, \lambda, \alpha, \mathbf{\beta}, \varepsilon)$ is the probability of community development score quantification. We use $\Pr(\mathbf{x}| \mathbf{z}^*)$ as the probability of generative process for the observed categorical value based on development score. And  $\Pr(\varepsilon)$ is as the probability of the error.

In the first  part of Eq. (1), we take  the  communities with shared entities  as association linked. It corresponds to the transformation of  community association network $\mathbf{Y}\in {\{0,1\}^{ n \times n}}$ from entity affiliation network $\mathbf{H}\in {\{0,1\}^{ n \times m}}$. That is, we set $\mathbf{Y}_{ij} = 1$ when ${  \mathbf{H}_{i,}} \cdot \mathbf{H}_{_{j,}}^T > 0$, and set $ {\mathbf{Y}_{ij}} = 0$ when ${  \mathbf{H}_{i,}} \cdot \mathbf{H}_{_{j,}}^T = 0$. Since community association network $\mathbf{Y}$\footnote{For simplicity, we set Y to be binary-valued. Note that it is not necessary to specify the number of entities, since two companies share at most one director.} is deterministically measured, the probability $\Pr(\mathbf{Y}|\mathbf{H})$ is 1.
Note that this probability is constant and introduced for comprehension.

 In the second part of Eq. (1), we characterize the network information as the social capital.  We assume the development score of a  community will benefit from its neighbors through information diffusion.  To be specific,  the development strength value $\mathbf{z}^{k+1}$ at time $k+1$ should  not only be determined  by development strength $\mathbf{z}^k$ at time $k$ and attribute $\mathbf{A}$ of their own, but also affected by the development strength of the connected communities. We formulate this as

\begin{small}
\begin{equation}\label{equ:2}
\mathbf{z}^{k+1} = \lambda \mathbf{Y}\mathbf{z}^{k} + \mathbf{A{}\beta}  + \mathbf{l}\alpha  + \mathbf{\varepsilon},
\end{equation}
\end{small}

where matrix $\mathbf{Y}$ is the  community association network, $\mathbf{A}$ is attribute information, $\lambda$ is the influence weight of aggregated links, $\beta$ is the coefficient parameter, $\alpha$ represents the fixed effect, $\mathbf{l}$  is n-dimensional vector of ones,  $\varepsilon \sim N(0,{\sigma_\varepsilon  ^2}\mathbf{I})$ is the error disturbance vector of the multivariate normal /gaussian distribution.

\begin{theorem}\label{cauchy}
The model in Eq. (2) has unique equilibrium and the community development score will converge to it.
\end{theorem}

As in  Eq. $\eqref{equ:2}$, we refer to the equilibrium as  $\mathbf{z}^{*} = \lambda \mathbf{Y}\mathbf{z}^{*} + \mathbf{A{}\beta}  + \mathbf{l}\alpha  + \mathbf{\varepsilon}$. Intuitively, the network effect is weak and parameter $\lambda$ is assumed smaller than the norm of the inverse of the  largest eigenvalue of $\mathbf{Y}$. Under such assumption, the matrix $\mathbf{I} - \lambda \mathbf{Y} \in {\mathbb{R}^{n \times n}}$ is non-singular and invertible. And thus, the equilibrium $\mathbf{z}^*$  is unique and specified as:

\begin{equation}\label{equ:3}
{\mathbf{z}^*} = {(\mathbf{I} - \lambda \mathbf{Y})^{ - 1}}(\mathbf{A{}\beta} + \mathbf{l}\alpha  + \mathbf{\varepsilon})
\end{equation}

Let ${\mathbf{z}^k} = {\mathbf{z}^*} + {\varepsilon ^k}$ and $\varepsilon ^{k + 1} = \lambda \mathbf{Y}{\varepsilon ^ k}$. Under the assumption that $\lambda$ is smaller than the norm of the inverse of the  largest eigenvalue of $\mathbf{Y}$, it is sufficient to show that  $\varepsilon ^{k + 1} = \lambda \mathbf{Y}{\varepsilon ^k} < \varepsilon ^ k$.
Therefore, the error converges to zero as $k$ grows and the equilibrium of $\mathbf{z}$ will converge to ${\mathbf{z}^*}$.

 From Theorem 1, we can deduce that the probability  $\Pr (\mathbf{z}^*|\mathbf{Y}, \mathbf{A}, \alpha, \mathbf{\beta}, \lambda, \varepsilon)$  is 1. It is the same with the first part of Eq. (1), since the score $\mathbf{z}^*$ is deterministically quantified as in  Eq. $\eqref{equ:3}$.

Also displayed in Figure 4(e), for the third part of Eq. (1), we assume the observed development category is generated from the unobserved development strength by  the logistic function; for the fourth part of Eq. (1), the error term $\varepsilon$  is supposed to follow the gaussian distribution, with parameter $\sigma _\varepsilon$ as error variance. Combining  these four parts together, we can derive the following from Eq. $\eqref{equ:1}$:

\begin{small}
\begin{equation}\label{equ:4}
\begin{array}{l}
 \Pr(\mathbf{H},  \mathbf{A}, \mathbf{x}; \mathbf{z}^*, \mathbf{Y}, \mathbf{\varepsilon}) \propto  \prod\limits_i {
 \frac{{\exp ( - (a\mathbf{z}_i^* + b)(1 - {\mathbf{x}_i}))}}{{1{\rm{ + }}\exp ( - (a\mathbf{z}_i^* + b))}} \cdot \frac{{\exp ( - {\varepsilon _i}^2/2{\sigma _\varepsilon }^2)}}{{{{(2\pi )}^{1/2}}{\sigma _\varepsilon }}}}\\

\\
s.t. \quad {\mathbf{z}^*} = {(\mathbf{I} - \lambda \mathbf{Y})^{ - 1}}(\mathbf{A{}\beta} + \mathbf{l}\alpha  + \mathbf{\varepsilon})
\end{array}
\end{equation}
\end{small}

In real applications, online social network also benefits information diffusion. As is shown in Figure 4(g), multi-relational social network between communities exist. To begin with,  we  aggregate the multiple relations  of community social network $\mathbf{F}$ as:

\begin{small}
\begin{equation}\label{equ:9}
\begin{array}{l}
 {\mathbf{S}_{ij}} = \sum\nolimits_q {{\xi _q}{\mathbf{H}_{i,}}{\mathbf{F}^{(q)}}\mathbf{H}_{j,}^T} \quad(\sum\nolimits_q {{\xi _q} = 1;} \ {\xi _q} > 0),

 \end{array}
\end{equation}
\end{small}

where   $\mathbf{\xi}$ is the weighting vector with $q$ dimensions. ${\mathbf{H}_{i,}}$ represents the members of  community $i$, ${\mathbf{F}^{(q)}}$ expresses the entity social networks with link type $q$. As in Figure 4(f), to relate multi-relational social links to  community development quantification, the joint probabilistic distribution is:

\begin{small}
\begin{equation}\label{equ:8}
\begin{array}{l}
 \Pr(\mathbf{F}, \mathbf{H},  \mathbf{A}, \mathbf{x}; \mathbf{z}^*, \mathbf{Y}, \mathbf{\xi}) \\
 \\

\propto  \Pr(\mathbf{S}|\mathbf{F},  \mathbf{H}, \mathbf{\xi}) \cdot  \Pr(\mathbf{Y}|\mathbf{H}) \cdot \Pr(\mathbf{x}| \mathbf{z}^*, a, b)\cdot  \Pr(\varepsilon) \cdot   \Pr(\mathbf{z}^*,  \mathbf{Y}| \mathbf{A}, \mathbf{S},  \lambda, \alpha, \mathbf{\beta}, \varepsilon, \eta,  w, \rho), \\
  \end{array}
\end{equation}
\end{small}

where $\Pr(\mathbf{S}| \mathbf{F}, \mathbf{H}, \mathbf{\xi})$ is the probability of community social network transformation process. Since community social network $\mathbf{S}$ is deterministically measured as in $\eqref{equ:9}$, the probability $\Pr(\mathbf{S}|\mathbf{F}, \mathbf{H}, \mathbf{\xi})$ is 1.  To leverage online social network information, we incorporate the social link effect $\lambda_2 \mathbf{S}\mathbf{z}^{*}$ to the community development score $\mathbf{z}^*$.

The development score can be represented as ${z^{k + 1}} = {\lambda _1}Y{z^k}{\rm{ + }}{\lambda _2}S{z^k}{\rm{ + A}}\beta  + {\rm{l}}\alpha  + \varepsilon $.  And thus, compared with Eq. (4), the learning target of the proposed model is:

\begin{small}

\begin{equation}
\begin{array}{l}
\Pr (F,H,A,x;{z^*},Y,\varepsilon ) \propto  \prod\limits_i {
 \underbrace{\frac{{\exp ( - {\varepsilon _i}^2/2{\sigma _\varepsilon }^2)}}{{{{(2\pi )}^{1/2}}{\sigma _\varepsilon }}}}_{\Pr(\varepsilon_i) }\cdot  \underbrace{\frac{{\exp ( - (a\mathbf{z}_i^* + b)(1 - {\mathbf{x}_i}))}}{{1{\rm{ + }}\exp ( - (a\mathbf{z}_i^* + b))}}}_{\Pr(\mathbf{x}_i| \mathbf{z}_i^*, a, b) }}\\

\\

where  \qquad \quad {\mathbf{z}^*} = {(\mathbf{I} - \lambda_1 \mathbf{Y} - \lambda_2 \mathbf{S} )^{ - 1}}(\mathbf{A{}\beta} + \mathbf{l}\alpha  + \mathbf{\varepsilon})\\

 \\
\qquad \quad \qquad \quad {\mathbf{S}_{ij}} = \sum\nolimits_q {{\xi _q}{\mathbf{H}_{i,}}{\mathbf{F}^{(q)}}\mathbf{H}_{j,}^T}  \quad(\sum\nolimits_q {{\xi _q} = 1;} \ {\xi _q} > 0)\\

\\

 \end{array}
\end{equation}

\end{small}

For parameter estimation, we
turn to the Newton Raphson Method.


\subsection{Network Formation Modeling}

While community development is affected by network diffusion, the development  can also affect network formation.  We call it selection bias, ignoring which would amplify the estimated network influence in community development strength and harm the corresponding quantification results.  As in Figure 4(g),  relating the selection process to quantify community development strength $\mathbf{z}^*$, we \emph{assume} the joint process of community development  and network formation follows a two-step sequential game: (1) community develops under the given network structure, (2) network links form and dissolve based on network structure and community development strength. As time goes to infinity, the  process of community development and network formation is equivalent to the maximum utility of  network formation and equilibrium development score under the network (\cite{Hsieh2017Specification1,mas1995microeconomic}).

We define ${u_{i}(\mathbf{Y})}=   \sum\nolimits_j {\varphi _{i,j}}({\mathbf{A}, \mathbf{Y}}) \cdot \eta + \mathbf{\overrightarrow{z}}^*_i \cdot \rho    + \sum\nolimits_j {\psi _{i,j}}({\mathbf{\overrightarrow{z}}^*}) \cdot w$ as the \textbf{utility} of each community in network formation,  where $\varphi$ represents network statistics, such as popularity (the total interaction for a community) and transitive triads, as well as  attribute similarity;  $\psi_{i,j}(\mathbf{\overrightarrow{z}}^*)=|\mathbf{\overrightarrow{z}}^*_i-\mathbf{\overrightarrow{z}}^*_j|$ is to quantify the possible cost of network formation; $\eta$, $w$ and $\rho$  are coefficient parameters. Meanwhile, the probability of development score $\mathbf{\overrightarrow{z}}^*$ is 1 given $\varepsilon$ since it is deterministically determined.  By modeling the whole network as a random variable and incorporating the community development strength,  the joint probability of network formation and community development is  as follows:

\begin{small}
\begin{equation}\label{equ:7}
\begin{array}{l}
 \Pr( \mathbf{Y}| \mathbf{\overrightarrow{z}}^*, \eta,  w, \rho) = \frac{{\exp (\sum\nolimits_i {{u_i}}(\mathbf{Y}) )}}{{\sum\nolimits_\mathbf{\bar{Y}} {\exp (\sum\nolimits_i {{u_i}}(\bar{\mathbf{Y}}) )} }}\\
\\


where. \quad {u_{i}(\mathbf{Y})}=   \sum\nolimits_j {\varphi _{i,j}}({\mathbf{A}, \mathbf{Y}}) \cdot \eta + \mathbf{z}^*_i \cdot \rho    + \sum\nolimits_j {\psi _{i,j}}({\mathbf{z}^*}) \cdot w\\


\end{array}
\end{equation}
\end{small}

It coincides with exponential random graph model \footnote{Each interaction is generated with bernoulli probability given attribute and network information. It coincides with exponential graph when modeling  graph formation (\cite{robins2004small}). }, where $\bar {\mathbf{Y}}$ is the  possible network.    Note  that $\mathbf{\varepsilon}$ exists in both community development strength quantification and network  formation modeling, which  joins the two process together and can then correct possible selection bias caused by network formation in community development strength quantification. The objective function can be then formalized as follows:

\begin{small}
\begin{equation}
\begin{array}{l}
 \mathop {\max }\limits_{{{\overrightarrow z}^*},\eta ,\rho ,w} L(O,{{\overrightarrow z}^*},\eta ,\rho ,w) \propto \underbrace {\frac{{\exp (\sum\nolimits_i {{u_i}(Y)} )}}{{\sum\nolimits_{\overline Y } {\exp (\sum\nolimits_i {{u_i}(\overline Y )} )} }}}_{\Pr (Y|,)}\prod\limits_i {\underbrace {\frac{{\exp ( - \overrightarrow \varepsilon {{_i^*}^2}/2\sigma _{\overrightarrow \varepsilon *}^2)}}{{{{(2\pi )}^{1/2}}{\sigma _{\overrightarrow \varepsilon *}}}}}_{\Pr (\overrightarrow \varepsilon *)}}  \\
  \\
 s.t.{\rm{  }}{u_i}(Y) = \sum\nolimits_j {{\varphi _{i,j}}(A,Y) \cdot } \eta  + \overrightarrow{z}_i^* \cdot \rho  + \sum\nolimits_j {{\psi _{i,j}}({{\overrightarrow z}^*}) \cdot } w \\
  \\
 {\rm{      }}{{\overrightarrow \varepsilon }^*} = {\rm{ }}{z^*} - {{ {\overrightarrow{z}}}^*} \\
\end{array}
\end{equation}
\end{small}

For parameter estimation, we
turn to the Bayesian estimation with the Metropolis-Hastings (M-H) algorithm.

%
%
%
%
%
%
%
%
%
%
%
%
%
%
%
%
%
\section{Experiment}

%
%
%

In this section, we apply the proposed
 model to  board and director network.  All experiments are conducted on machines with Intel(R) Xeon(R) CPU of 2.5GHz and 32G memory. We use python for data preprocessing and  Fortran for  parameter estimations.

\subsection{Experimental Settings}

We conduct experiments on both real-world data and synthetic data.

\textbf{Real Dataset:} The data comprises of directors and corporations information collected by BoardEx.  It includes two aspects: (1) the corporation centered information containing basic financial properties; (2) the director oriented information comprising of basic attributes, membership relations.  The data set  consists of 8,186 corporations and 76,362 directors ranging from the  year 1990 to the year 2009.  The  board and director affiliation network can be extracted\footnote{For convenience, we cluster the communities in the economic graph into 35 super-communities}. For multi-relational social network between directors, we select four types: the NFP (Not For Profit) association, unlisted association, education association and other association.  Finally, we derive an economic graph with 20 attributes\footnote{TimeRetirement, TimeRole, TimeBrd, TimeInCo, AvgTimeOthCo, TotNoUnLstdBrd, TotNoLstdBrd, TotNoOthLstdBrd, NoQuals, Succession(Tenure), NumberDirectors, Attrition, GenderRatio, NationalityMix, STDEVTimeBrd, STDEVTimeInCo, STDEVTotNoLstdBrd, STDEVTotCurrNoLstdBrd, STDEVNoQuals, STDEVAge(Age Deviation)}, 4 social network and 1 affiliation network.

For community predicted development categorization information,  we assign the positive label to communities with higher salary (larger employee size) and negative vice versa correspondingly. Following the fact that  categorical value is based on analyst's prediction (rising and declining) with average prediction accuracy  above 92\% (\cite{Hong2003Analyzing}), the approximate substitution of mean real value based rating (positive/ negative) is acceptable. Meanwhile,   the real value of salary (employee size) is taken as the ground truth of community development strength scoring.

%
%
%
%
%
%
%
%
%
%
%
%
%
\textbf{Synthetic Dataset:} We generate synthetic data similar to the real data. That is, we generated the synthetic dataset with the same  settings of board and director dataset. We vary and amplify the dataset by expanding its size based on the real dataset.  The number of communities ranges from 896 to 137,760,  and attribute dimension is from 20 to 100.


%

%
%
%
%

\textbf{Evaluation Perspectives:} We evaluate the proposed model from  three perspectives: (1) factor contribution, (2) model accuracy, (3) model scalability and simulation. To remove the high-percentage of missing information as well as reduce the variance, the average results from the  year 2000 to 2009 are taken for evaluation.

\subsection{ Model Accuracy}
Note that we have proposed a new problem and no direct methods are specifically designed to solve our proposed problem, we choose classic methods\footnote{To the best of our knowledge, no recent proposed methods are suitable for this task due to the totally new settings.} that can provide intermediate results representing as real-valued score indicator for comparison. Generally, we slightly modify them to accommodate to our problem as baselines and classify them into three types: (1) utilizing network as diffusion channels (\textbf{LNP}); (2) ignoring network effects (\textbf{Logistic Regression}); (3) extracting deep network structure features (\textbf{LSM}, \textbf{DeepWalk}). The detailed information is as follows:
 \begin{itemize}

\item \textbf{LNP:} \cite{Wang2008Label} assumes that each data point can be linearly reconstructed from its neighborhood, which can propagate the labels from the labeled points to the whole data set using these linear neighborhoods. It is nearly equivalent to weak-supervised ranking with categorical information setting as in \cite{Gao2012Semi}, and our proposed method belongs to this type. It's a relaxed version of network based development strength quantification. The affiliation network and social network are combined as community network with the learnt parameter as in Table 2.

\item \textbf{Logistic Regression:} We simply ignore network effects for classification, and the loss function score along training process is taken for development score.

 \item \textbf{LSM:} \cite{Hoff2002Latent} assumes the snapshot of a static social network is generated based on the positions of individuals in an unobserved social space, a space of unobserved latent characteristics representing  transitive tendencies in network relations. The training process can be taken as leveraging network structures for development strength quantification. Also, the network is specified as the same with \textbf{LNP}.

 \item \textbf{DeepWalk:} \cite{Perozzi2014DeepWalk} proposes to learn latent representations of vertices in a network. These latent representations
encode social relations in a continuous vector space. It allowed a global view of the network information for development strength quantification. Also, the network is specified as the same with \textbf{LNP} and \textbf{LSM}.

\end{itemize}
\textbf{Accuracy Evaluation:} For accuracy evaluation,  we compare the intersection dissimilarity between  the learnt  score and the ground truth.  The intersection dissimilarity measure ($\mathbf{IDM}$) (\cite{Boldi2005TotalRank}) is
to measure the dissimilarity ranking between ground truth and the estimated value  as$\mathbf{IDM }= 1/2n\sum\nolimits_{j = 1}^n {|{\mathbf{z}_j}\Delta {\bar{\mathbf{z}}_j}|/2j}$, where $\mathbf{z}$ represents the development score estimated by
different methods, $\bar{\mathbf{z}}$ serves as the value of the ground truth, and $\Delta$
denotes the symmetric set difference. The larger the value of $\mathbf{IDM}$ is, the less similar $\mathbf{z}$ and $\bar{\mathbf{z}}$ are.

\section{Conclusion}

In this paper, we introduce a new problem of prospecting the development strength of communities. Accounting for network diffusion effects in community development, we refine  affiliation network and social network with attributes as economic graph. Specifically, we proposed a hybrid of discriminative and generative model for network based development score estimation. To account for selection bias, a  random graph generation model is jointly proposed. Moreover, we jointly integrate and weigh multi-relational network for development strength estimation. The substantive experimental results reveal the competitive advantage over the state-of-the art approaches. As for future work, we can combine economic network with other vital information, such as community profiling, to potentially enhance the proposed model. Meanwhile, the model can be extended where general categorical attribute (not just binary rating) are considered.

\bibliographystyle{splncs03}
\bibliography{firmperformance0731-dasfaa}


\end{document}